\documentclass[12pt,draftcls,onecolumn]{IEEEtran}
\usepackage{amsfonts}
\usepackage{latexsym}
\usepackage{newlfont}
\usepackage{graphicx}
\usepackage{mathtools}
\usepackage{booktabs}
\usepackage{tikz}
\usetikzlibrary{snakes}

\newcommand{\Ac}{\mathcal{A}}
\newcommand{\Ic}{\mathcal{I}}
\newcommand{\Nc}{\mathcal{N}}

\newcommand{\real}{\mathbb{R}}

\newcommand{\bn}{\mathbf{n}}
\newcommand{\bx}{\mathbf{x}}
\newcommand{\by}{\mathbf{y}}
\newcommand{\bw}{\mathbf{w}}
\newcommand{\bmu}{\boldsymbol{\mu}}
\newcommand{\bA}{\mathbf{A}}
\newcommand{\bC}{\mathbf{C}}
\newcommand{\ba}{\mathbf{a}}
\newcommand{\bc}{\mathbf{c}}
\newcommand{\be}{\mathbf{e}}
\newcommand{\bg}{\mathbf{g}}
\newcommand{\bB}{\mathbf{B}}
\newcommand{\bE}{\mathbf{E}}

\newcommand{\bD}{\mathbf{D}}
\newcommand{\bG}{\mathbf{G}}
\newcommand{\bH}{\mathbf{H}}

\newcommand{\bN}{\mathbf{N}}
\newcommand{\bP}{\mathbf{P}}
\newcommand{\bR}{\mathbf{R}}
\newcommand{\bI}{\mathbf{I}}

\newcommand{\bU}{\mathbf{U}}
\newcommand{\bV}{\mathbf{V}}
\newcommand{\bv}{\mathbf{v}}
\newcommand{\bz}{\mathbf{z}}

\newcommand{\bLam}{\mathbf{\Lambda}}

\newcommand{\bGamma}{\mathbf{\Gamma}}

\newcommand{\bzero}{\mathbf{0}}
\newcommand{\diag}{\textrm{Diag}}
\newcommand{\EX}{\mathbb E}
\newcommand{\bdiag}{\textrm{block-Diag}}

\newcommand{\comment}[1]{}
\newcommand{\tr}{{\text{tr}\,}}

\newtheorem{thm}{Theorem}

\newtheorem{lem}[thm]{Lemma}

\newtheorem{rem}[thm]{Remark}

\usepackage{ifpdf}

\usepackage{cite}
\usepackage{array}

\usepackage{mdwmath}
\usepackage{mdwtab}

\usepackage{eqparbox}
\usepackage{url}

\begin{document}
\sloppy
%
\title{Greedy Adaptive Compression in Signal-Plus-Noise Models}
%
%
%

\author{Entao~Liu, \IEEEmembership{Member,~IEEE},
        Edwin~K.~P.~Chong, \IEEEmembership{Fellow,~IEEE},
        and~Louis~L.~Scharf, \IEEEmembership{Life Fellow,~IEEE}
\thanks{This work was supported in part by DARPA contract
N66001-11-C-4023, ONR contract N00014-08-1-110, NSF grant CFF-1018472, and AFOSR
contract FA 9550-10-1-0241. This paper was presented in part at CISS 2012.}
\thanks{Entao Liu is with the Department of ECE, Colorado State University. e-mail: (entao.liu@colostate.edu).}
\thanks{Edwin K. P. Chong is with the Department of ECE and Department
of Mathematics, Colorado State University. e-mail: (edwin.chong@colostate.edu).}
\thanks{Louis L. Scharf is with the Departments of Mathematics and Statistics, Colorado State University. e-mail: (scharf@engr.colostate.edu ).
}
}
\maketitle

\begin{abstract}
The purpose of this article is to examine greedy adaptive
measurement policies in the context of a linear Gaussian measurement
model with an optimization criterion based on information gain.
In the special case of sequential scalar measurements,
we provide sufficient conditions under which the greedy policy actually is optimal in the sense of maximizing the net information gain.
We also discuss cases where the greedy policy is provably not optimal.

\end{abstract}


%
\IEEEpeerreviewmaketitle

\section{Introduction}

\IEEEPARstart{C}{onsider} a signal of interest $\bx$, which is a random vector taking values in $\real^N$ with (prior) distribution
$\Nc(\bmu,\bP_0)$ (i.e., $\bx$ is Gaussian distributed with mean $\bmu$ and $N\times N$ covariance matrix $\bP_0$). The signal $\bx$
is carried over a noisy channel to a sensor, according to the model
$\bz:=\bH\bx+\bn$ where $\bH \in \real^{K \times N}$ is a full rank
channel matrix. For simplicity, in this paper we focus on the case
where $K\ge N$, though analogous results are obtained when $K< N$. The problem is to compress $m$ realizations of $\bz$ ($\bz_k=\bH\bx+\bn_k$, $k=1,\ldots,m$)
with $m$ measurements (where $m$ is specified upfront). But the implementation of each compression has a noise penalty. So, the $k$th compressed measurement is
\begin{equation}\label{on}
\by_k = \bA_k(\bH\bx+\bn_k) + \bw_k
\end{equation}
where the compression matrix $\bA_k$ is $L\times K$. Consequently, the measurement $\by_k$ takes values in $\real^L$.
Assume that the measurement noise $\bw_k\in \real^L$ has distribution
$\Nc(\bzero,\bR_{ww})$ and channel noise $\bn_k\in \real^K$ has
distribution $\Nc(\bzero,\bR_{nn})$. The measurement and channel
noise sequences 
are independent over $k$ and independent of each other.
Equivalently, we can rewrite (\ref{on}) as
\begin{equation}
\by_k = \bA_k\bH\bx+(\bA_k\bn_k + \bw_k)
\end{equation}
and consider $\bA_k\bn_k + \bw_k$ as the total noise with distribution $\Nc(\bzero,\bA_k\bR_{nn}\bA^T_k+\bR_{ww})$.


We consider the following adaptive (sequential) compression problem.
For each $k=1,\ldots,m$, we are allowed to choose the
compression matrix $\bA_k$ (possibly subject to some constraint).
Moreover, our choice is allowed to depend on the entire history of
measurements up to that point:
$\Ic_{k-1} = \{\by_1,\ldots,\by_{k-1}\}$.

Let the posterior distribution of $\bx$
given $\Ic_k$ be $\Nc(\bx_k,\bP_k)$. More
specifically, $\bP_k$ can be written recursively for $k=1,\ldots,m$ as
\begin{equation}\label{PP0}
\bP_k = \bP_{k-1} - \bP_{k-1}\bB_k^T(\bB_k\bP_{k-1}\bB_k^T +\bN_k)^{-1}\bB_k\bP_{k-1},
\end{equation}
where $\bB_k:=\bA_k\bH$ and $\bN_k:=\bA_k\bR_{nn}\bA_k^T+\bR_{ww}$.
If this expression seems a little unwieldy, by the Woodbury identity a simpler version is
\begin{equation}\label{PP}
\bP_k = \left(\bP_{k-1}^{-1} + \bB_k^T\bN_k^{-1}\bB_k\right)^{-1},
\end{equation}
assuming that $\bP_{k-1}$ and $\bN_k$ are nonsingular.
Also define the \emph{entropy} of the posterior distribution of
$\bx$ given $\Ic_k$:
\begin{equation}\label{HH}
H_k = \frac{1}{2}\log\det(\bP_k) + \frac{N}{2}\log(2\pi e).
\end{equation}
The first term $\det(\bP_k)$ is actually proportional to the volume of the error concentration ellipse for $\bx-\EX[\bx|\Ic_k]$.

We focus on a common information-theoretic criterion for choosing
the compression matrices: for the $k$th compression matrix, we pick $\bA_k$ to
maximize the \emph{per-stage information gain}, defined as $H_{k-1} - H_k$.
For reasons that will be made clear later, we refer to this strategy
as a \emph{greedy} policy.
The term \emph{policy} simply refers to a rule for picking $\bA_k$ for each
$k$ based on $\Ic_{k-1}$.

Suppose that the overall goal is
to maximize the \emph{net information gain}, defined as $H_0 - H_m$.
We ask the following questions:
Does the greedy policy achieve this goal?
If not, then what policy achieves it?
How much better is such a policy than the greedy one?
Are there cases where the greedy policy does achieve this goal?
In Section~\ref{gp}, we analyze the greedy policy and compute its
net information gain. In Section~\ref{op}, to find the
net information gain of the optimal policy, we introduce a relaxed
optimization problem, which can be solved as a
 water-filling problem. In Section~\ref{g=o}, we derive two
sufficient conditions under which the greedy policy is optimal.
In Section~\ref{examples}, we give examples for which the greedy
policy is not optimal.

\section{Greedy Policy}\label{gp}
\subsection{Preliminaries}

We now explore how the {\it greedy policy} performs for the adaptive
measurement problem. Before proceeding, we first make some remarks on the information gain
criterion:
\begin{itemize}
\item Information gain as defined in this paper also goes by the name \emph{mutual
information} between $\bx$ and $\by_k$ in the case of per-stage
information gain, and between $\bx$ and $\Ic_m$ in the
case of net information gain.
\item The net information gain can be written as the cumulative sum
of the per-stage information gains:
\begin{equation*}
H_0 - H_m = \sum_{k=1}^m (H_{k-1}-H_k).
\end{equation*}
This is why the greedy policy is named as such; at each stage $k$,
the greedy policy simply maximizes the immediate (short-term) contribution
$H_{k-1}-H_k$ to the overall cumulative sum.
\item Using the formulas (\ref{PP0}) and (\ref{HH}) for $H_k$ and $\bP_k$, we can write
\begin{align}\label{IG}
&H_{k-1} - H_k= -\frac{1}{2}\log\det\big(\bI_N \nonumber\\
 &  -\bP_{k-1}\bB_k^T(\bB_k\bP_{k-1}\bB_k^T +\bN_k)^{-1}\bB_k\big)
\end{align}
 where $\bI_N$ is the $N\times N$ identity matrix.
In other words, at the $k$th stage, the greedy policy minimizes (with respect to
$\bA_k$)
\begin{equation}
\log\det\left(\bI_N - \bP_{k-1}\bB_k^T(\bB_k\bP_{k-1}\bB_k^T
+\bN_k)^{-1}\bB_k\right).
\label{eqn:ig}
\end{equation}
\item Equivalently, using the other formula (\ref{PP}) for $\bP_k$, the greedy
policy maximizes
\begin{equation}
\log\det\left(\bP_{k-1}^{-1} + \bB_k^T\bN_k^{-1}\bB_k\right)
\end{equation}
at each stage.
For the purpose of optimization, the $\log$ function in the
objective functions above can be dropped, owing to its
monotonicity.
\end{itemize}

It is worth noting that we may dispense with the
assumption of Gaussian distributed variables and argue that we are simply minimizing
$\det\bP_k$, which is proportional to the volume of the error concentration ellipse defined by $(\bx-\hat{\bx}_{k-1})^T\bP^{-1}_k(\bx-\hat{\bx}_{k-1})\le 1$.
Notice that the greedy policy does not use the values of
$\by_1,\ldots,\by_{k-1}$; its choice of $\bA_k$ depends only on
$\bP_{k-1}$, $\bR_{nn}$ and $\bR_{ww}$. In fact, the formulas above
show that information gain is a deterministic function of the
model matrices (in our particular setup).
This implies that the optimal policy can
be computed by deterministic dynamic programming. In general, we
would not expect the greedy policy to solve such a dynamic
programming problem. However, as we will see in following sections, there are cases
where it does.

\subsection{Sequential Scalar Measurements}

This subsection is devoted to the special case where $L=1$ (i.e., each measurement is a scalar). Accordingly,
we can write
$\bA_k = \ba_k^T$, where $\ba_k\in\real^K$, $\bR_{ww}=\sigma_w^2$,
and $\bR_{nn}=\sigma_n^2\bI_K$. Accordingly, the scalar measurement
$y_k$ is given by
\begin{equation}\label{ssm}
y_k=\ba_k^T(\bH\bx+\bn_k)+w_k,
\end{equation}
for $k=1,\ldots,m$. This problem is the problem of designing the columns of compression matrix $\bA=[\ba_1,\ldots,\ba_m]$ sequentially, one
at a time. In the special case $\bn_k=\bzero$, the measurement model is
\begin{equation}\label{ssm1}
\by=\bA^T \bH\bx+\bw,
\end{equation}
where $\by\in \real^m$ is called the measurement vector, and $\bw$ is a white Gaussian noise vector. In this context, the construction of a ``good'' compression matrix $\bA$ to convey information about $\bx$
is also a topic of interest. When $\by=\bA^T \bx+\bw$, this is a problem of greedy adaptive noisy compressive sensing. Our solution is a more general solution than this for the more general problem (\ref{ssm1}). In this more general problem, the uncompressed measurement $\bH\bx+\bn_k$
is a noisy version of the filtered state $\bH\bx$, and compression by $\ba_k$ introduces measurement noise $w_k$ and colors the channel noise $\bn_k$.

The concept of sequential scalar measurements in a closed-loop
fashion has been discussed in a number of recent papers; e.g., \cite{Ela07, JXL, JiD09, CaH08, HaC10, HCN10, Car, LC, LCS, AHN, KAN}. The objective function for the optimization here can take a number of possible forms, besides the net information gain. For example, in \cite{JXL}, the objective is to maximize the posterior variance of the expected measurement.

If the $\ba_k$ can only be chosen from
a prescribed \emph{finite} set, the optimal design of $\bA$ is essentially a sensor selection problem (see \cite{JoBo},\cite{REJVBBP}), where the greedy policy has been shown to perform well. For example, in the problem of sensor selection under a submodular objective function subject to a uniform matroid constraint
\cite{SBV},
the greedy policy is suboptimal with a provable bound on its performance, using bounds from optimization of submodularity functions \cite{NW},\cite{CCPV}.

Consider a constraint of the form $\|\ba_k\|\le 1$ for
$k=1,\ldots,m$ (where $\|\cdot\|$ is the Euclidean norm in $\real^K$), which is much more relaxed than a prescribed finite set. The constraint that $\bA$ has unit-norm columns is a standard setting for compressive sensing \cite{Don}. The expression in (\ref{eqn:ig}) simplifies to
\begin{equation}\label{IG1}
\log\det\left(\bI_N - \frac{\bP_{k-1}\bH^T\ba_k\ba_k^T\bH}
{\ba_k^T\bH\bP_{k-1}\bH^T\ba_k + \sigma_n^2\|\ba_k\|^2+\sigma_w^2}
\right).
\end{equation}
This expression further reduces (see \cite[Lemma~1.1]{DZ}) to
\begin{equation}\label{IG2}
\log\left(1-\frac{\ba_k^T \bH\bP_{k-1}\bH^T\ba_k}{\ba_k^T \bH\bP_{k-1}\bH^T\ba_k+\sigma_n^2\|\ba_k\|^2+\sigma_w^2}\right).
\end{equation}
Combining (\ref{IG}) and (\ref{IG2}), the information gain at the $k$th step is
\begin{align}\label{infogain}
\MoveEqLeft H_{k-1}-H_k\nonumber\\
 &=-\frac{1}{2}\log \left(1 - \frac{1}{1 + (\sigma_n^2\|\ba_k\|^2+\sigma_w^2)/ \ba_k^T \bH \bP_{k-1} \bH^T \ba_k}\right).
\end{align}
It is obvious that the greedy policy maximizes
\begin{equation}\label{ratio}
\frac{\ba_k^T \bH\bP_{k-1}\bH^T\ba_k}{\sigma_n^2\|\ba_k\|^2+\sigma_w^2}
\end{equation}
to obtain the maximal information gain in the $k$th
step. Clearly, the measurement $y_k$ may be written as
\begin{align}
y_k&=\ba_k^T \left(\bH\hat{\bx}_{k-1}+\bH(\bx-\hat{\bx}_{k-1})+\bn_k\right)+w_k\nonumber\\
   &=\ba_k^T \bH\hat{\bx}_{k-1}+\ba_k^T \bH(\bx-\hat{\bx}_{k-1})+\ba_k^T \bn_k+w_k.
\end{align}
 Then (\ref{ratio}) is simply the ratio of variance components: the numerator is $\EX \ba_k^T\bH (\bx-\hat{\bx}_{k-1})(\bx-\hat{\bx}_{k-1})^T \bH^T \ba_k$,
$\hat{\bx}_{k-1}=\EX [\bx|\Ic_{k-1}]$, and the denominator is
$\EX(\ba_k^T\bn_k+w_k)^2$. So the goal for the greedy policy is to select $\ba_k$ to maximize signal-to-noise ratio, where the signal is taken
to be the part of the measurement $y_k$ that is due to error $\bx-\hat{\bx}_{k-1}$ in the state estimate and noise is taken to be the sum of $\ba_k^T\bn_k$ and $w_k$. This is reasonable,
as $\hat{\bx}_{k-1}$ is now fixed by $\{y_1,\ldots,y_{k-1}\}$, and only variance components can be controlled by the measurement vector $\ba_k$.

The greedy policy can be described succinctly in terms of certain
eigenvectors, as follows.
Denote the eigenvalues of $\bD_k:=\bH\bP_{k}\bH^T$ by $\lambda^{(k)}_1 \ge \lambda^{(k)}_2 \ge \cdots \ge \lambda^{(k)}_N\ge\lambda^{(k)}_{N+1}=\cdots=\lambda^{(k)}_K=0 $. For simplicity,
when $k=0$ we may omit the superscript and write $\lambda_i:=\lambda_i^{(0)}$ for $i=1,\ldots,K$.
Since $\bP_0$ is a covariance matrix, which is symmetric, $\bD_0$ is
also symmetric, and there exist corresponding
orthonormal eigenvectors $\{\bv_1, \bv_2, \ldots, \bv_K\}$. Clearly,
\begin{equation}
\frac{\ba_1^T \bD_{0}\ba_1}{\sigma_n^2\|\ba_1\|^2+\sigma_w^2}\le \frac{\lambda_1\|\ba_1\|^2}{\sigma_n^2\|\ba_1\|^2+\sigma^2_w}
\le \frac{\lambda_1}{\sigma_n^2+\sigma^2_w}.
\end{equation}
The equalities hold when $\ba_1$ equals $\bv_1$, which is the eigenvector of $\bD_0$ corresponding to its largest eigenvalue
$\lambda_1$; we take this to be what the greedy policy picks. If eigenvalues are repeated, we simply pick the eigenvector with smallest index $i$. After picking $\ba_1=\bv_1$, by (\ref{PP0}) we have
\begin{align}
\bP_1=\bP_0-\frac{\bP_0\bH^T\bv_1\bv_1^T \bH \bP_0}{\sigma^2+\lambda_1}
\end{align}
where $\sigma^2:=\sigma_n^2+\sigma_w^2$. We can verify the following:
\begin{align}\label{v1}
\bD_1\bv_i&=\Big(\bH\big(\bP_0-\frac{\bP_0\bH^T\bv_1\bv_1^T \bH \bP_0}{\sigma^2+\lambda_1}\big)\bH^T\Big)\bv_i \nonumber\\
     &=\Big(\bD_0-\frac{\bD_0\bv_1\bv_1^T\bD_0}{\sigma^2+\lambda_1}\Big)\bv_i\nonumber\\
     &=\Big(\bD_0-\frac{\lambda_1^2\bv_1\bv_1^T}{\sigma^2+\lambda_1}\Big)\bv_i\nonumber\\
     &=\lambda_i\bv_i ,\quad \text{for } i=2\ldots,K,
\end{align}
and
\begin{align}\label{v2}
\bD_1\bv_1&=\left(\bD_0-\frac{\lambda_1^2\bv_1\bv_1^T}{\sigma^2+\lambda_1}\right)\bv_1\nonumber\\
          &=\Big(\frac{1}{\lambda_1}+\frac{1}{\sigma^2} \Big)^{-1}\bv_1.
  \end{align}
So we see that $\bD_1$ has the same collection of eigenvectors as $\bD_0$ ,
and the nonzero eigenvalues of $\bD_1$ are $(1/\lambda_1+ 1/\sigma^2 )^{-1}, \lambda_2, \ldots, \lambda_N$. By induction,
 we conclude that, when applying the greedy policy, all the $\bD_k$s for $k=0,\ldots,m$ have the same collection of eigenvectors and the greedy policy always picks the compressors $\ba_k$, $k=1,\ldots,m$, from the set of eigenvectors
$\{\bv_1,\ldots,\bv_N\}$. The implication is that this basis for the invariant subspace $\langle \bV\rangle$ for the prior measurement covariance $\bD_0$ may be used to define a prescribed finite set of compression vectors from which compressors are to be drawn. The
greedy policy then amounts to selecting the compressor $\ba_k$ to be the eigenvector of $\bD_k$ with eigenvalue $\lambda^{(k)}_1$. In other words, the greedy policy simply re-sorts the eigenvectors of $\bD_0$, step-by-step, and selects the one with
maximum eigenvalue.

Consequently, after applying $m$ iterations of the greedy policy, the net information gain is
\begin{align}\label{greedy}
 \MoveEqLeft H_0-H_m=\sum_{k=1}^m \max_{\|\ba_k\|\le 1} (H_{k-1}-H_k)\nonumber\\
  &=-\frac{1}{2} \sum_{k=1}^m \log\left(\frac{\sigma^2}{\lambda^{(k-1)}_1+\sigma^2}\right) \nonumber\\
  &=\frac{1}{2}\log \prod_{k=1}^m \left(1+\frac{\lambda_1^{(k-1)}}{\sigma^2}\right)
\end{align}
where $\lambda_1^{(k-1)}$, the largest eigenvalue of $\bD_{k-1}$, is computed iteratively from the sequence $\bP_0,\ldots,\bP_{k-1}$.

\subsection{Example of the Greedy Policy}

Suppose that the uncompressed measurements are $\bz_k=\bx+\bn_k$, $k=1,\ldots,m$, with
$\bP_0=\lambda\bI_N$, indicating no prior indication of shape for
the error covariance matrix. Assume that
$\bR_{ww}=\sigma_w^2\bI_N$ and $\bR_{nn}=\sigma_n^2\bI_N$. The
choice of orthonormal eigenvectors for $\bD_0=\bP_0$ is arbitrary, with $\bV=\bE=[\be_1,\ldots,\be_N]$ (the standard basis for $\real^N$)
a particular choice that minimizes the complexity of compression. So compressed measurements will consist of the noisy measurements $y_k=\be^T_{(k)}\bz+w_k$.

 After picking $\ba_1=\be_1$, the eigenvalues of $\bP_1$ are $\lambda^{(1)}_1=\cdots=\lambda^{(1)}_{N-1}=\lambda$, $\lambda^{(1)}_N=(\frac{1}{\lambda}+\frac{1}{\sigma^2})^{-1}$. Analogously, after picking
 $\ba_2=\be_2$, the
eigenvalues of $\bP_2$ are $\lambda^{(2)}_1=\cdots=\lambda^{(2)}_{N-2}=\lambda$, $\lambda^{(2)}_{N-1}=\lambda^{(2)}_N=(\frac{1}{\lambda}+\frac{1}{\sigma^2})^{-1}$, and so on.
If $m\le N$, then after $m$ iterations of the greedy policy the eigenvalues of $\bD_m$ are
$\lambda^{(m)}_1=\cdots=\lambda^{(m)}_{N-m}=\lambda$,
$\lambda^{(m)}_{N-m+1}=\cdots=\lambda^{(m)}_N=(\frac{1}{\lambda}+\frac{1}{\sigma^2})^{-1}$.
In the first $m$ iterations, the per-step information gain
is $\frac{1}{2}\log(1+\lambda/\sigma^2)$.

If $m>N$, after $N$ iterations of the
greedy policy,
$\lambda^{(N)}_1=\cdots=\lambda^{(N)}_N=(\frac{1}{\lambda}+\frac{1}{\sigma^2})^{-1}$.
We now simply encounter a similar situation as in the very
beginning. We update $\lambda\leftarrow
(\frac{1}{\lambda}+\frac{1}{\sigma^2})^{-1}$ and $m\leftarrow
(m-N)$. The analysis above then applies again, leading to a
round-robin selection of measurements.

\section{Optimal Policy and Relaxed Optimal Policy}\label{op}
\subsection{Optimal Policy}
In this subsection we consider the problem of maximizing the net
information gain, subject to the unit-norm constraint:
\begin{equation}\label{un}
\begin{split}
&\text{maximize      }  \sum_{k=1}^m (H_{k-1}-H_k),\\
&\text{subject to   } \|\ba_k\|\le 1, ~k=1,\ldots, m.
\end{split}
\end{equation}
The policy that maximizes (\ref{un}) is called the \emph{optimal
policy}.

The objective function can be written as
\begin{align}\label{IG3}
 \MoveEqLeft \sum_{k=1}^m (H_{k-1}-H_k)\nonumber\\
 &=-\frac{1}{2}\sum_{k=1}^m \log \frac{\det (\bP_{k})}{\det  (\bP_{k-1})}\nonumber\\
 &=\frac{1}{2}\log \frac{\det (\bP_0)}{\det (\bP_m)}\nonumber\\
&=\frac{1}{2}\log \det(\bP_0)\det\Big(\bP_0^{-1}+\sum_{k=1}^m \frac{\bH^T\ba_k \ba_k^T\bH}{\|\ba_k\|^2\sigma_n^2+\sigma_w^2}\Big)\nonumber\\
&=\frac{1}{2}\log \det (\bI_m+\bC^T\bD_0\bC )
\end{align}
where
\begin{equation}\label{ea}
\bC:=[\bc_1,\ldots,\bc_m]:=\left[\frac{\ba_1}{\sqrt{\|\ba_1\|^2\sigma_n^2+\sigma_w^2}},\ldots,\frac{\ba_m}{\sqrt{\|\ba_m\|^2\sigma_n^2+\sigma_w^2}}\right].
\end{equation}
Assume that the eigenvalue decomposition $\bD_0=\bV\bLam\bV^T$,
where $\bLam=\diag(\lambda_1,\lambda_2,\ldots,\lambda_K)$ and
$\bV=[\bv_1,\ldots,\bv_K]$. (The notation
$\diag(\lambda_1,\lambda_2,\ldots,\lambda_K)$ means the diagonal
matrix with diagonal entries $\lambda_1,\ldots,\lambda_K$.)
Then, continuing from (\ref{IG3}),
\begin{align}\label{IG4}
 \MoveEqLeft \sum_{k=1}^m (H_{k-1}-H_k)\nonumber\\
&=\frac{1}{2}\log \det\left(\bI_m+\bC^T\bV\bLam\bV^T\bC \right)\nonumber\\
&=\frac{1}{2}\log \det \left(\bI_m+\bG^T\bLam \bG \right)
\end{align}
where
\begin{equation}\label{ge}
\bG:=[\bg_1,\ldots,\bg_m]:=\bV^T \bC.
\end{equation}
 Since $\bV$ is nonsingular, the map $\bc_k \mapsto \bg_k=\bV^T \bc_k$ is one-to-one.

The constraint $\|\ba_k\|\le 1$ implies that
$\|\bg_k\|^2=\|\bc_k\|^2\le \sigma^{-2}$, so the constraint in
(\ref{un}) can be written as $(\bG^T\bG)_{ii}\le \sigma^{-2}$ for $i=1,\ldots,m$.
The problem (\ref{un}) is
actually equivalent to the maximum a posteriori probability (MAP)
problem (see \cite{BV} and \cite{SBV}).

\subsection{Relaxed Optimal Policy}

To help characterize the optimal policy (solution to (\ref{un})), we
now consider an alternative optimization problem with the same
objective function in (\ref{un}) but a relaxed constraint:
\begin{equation}\label{aun}
\begin{split}
&\text{maximize      }  \sum_{k=1}^m (H_{k-1}-H_k),\\
&\text{subject to   } \frac{1}{m}\sum_{k=1}^m\|\ba_k\|\le 1,
\end{split}
\end{equation}
i.e., the columns of $\bA$ have \emph{average} unit norm. We will
call the policy that maximizes (\ref{aun}) the \emph{relaxed optimal
policy}.

The average unit-norm constraint in (\ref{aun}) is equivalent to
$\tr \bG^T\bG= \sum_{k=1}^m
\|\bg_k\|^2 \leq \sigma^{-2}m$. With the scaling
\begin{equation}\label{gtg}
\widetilde{\bG}:=\sigma m^{-1/2}\bG,
\end{equation}
the constraint $\tr  \bG^T\bG \leq \sigma^{-2}m$ becomes
$\tr \widetilde{\bG}^T \widetilde{\bG}\leq 1$.
Hence, the relaxed optimization problem (\ref{aun}) is equivalent to
\begin{equation}\label{max}
\begin{split}
&\text{maximize      }  \frac{1}{2}\log  \det (\bI_m+\widetilde{\bG}^T\widetilde{\bLam}\widetilde{\bG}),\\
&\text{subject to   }\tr \widetilde{\bG}^T\widetilde{\bG}\le 1
\end{split}
\end{equation}
where $\widetilde{\bLam}=\diag(\Lambda_1,\ldots,\Lambda_N)$ and $\Lambda_i:=\frac{m\lambda_i}{\sigma^2}$, for $i=1,\ldots,N$.

To solve (\ref{max}), let us recall the following known results from \cite{Wit}.
\begin{lem}\label{wit_l1}
Given any $\lambda_1\ge \lambda_2 \ge \ldots \ge \lambda_q>0$, there
exists a unique integer $r$, with $1\le r \le q$, such that
for $1\le k\le r$ we have
\begin{equation}
\frac{1}{\lambda_k}<\frac{1}{k}\Big(1+\sum_{j=1}^k \frac{1}{\lambda_j}\Big),
\end{equation}
while for indices $k$, if any, satisfying $r<k\le q$ we have
\begin{equation}
\frac{1}{\lambda_k}\ge \frac{1}{k}\Big(1+\sum_{j=1}^k \frac{1}{\lambda_j}\Big).
\end{equation}
\end{lem}
\begin{lem}
For $\lambda_1\ge \lambda_2 \ge \ldots \ge \lambda_q>0$ and $r$ as in Lemma~\ref{wit_l1}, the sequence
\begin{equation}
M_k=\Big(\frac{1}{k}+\frac{1}{k}\sum_{j=1}^k \frac{1}{\lambda_j}\Big)^k \prod_{i=1}^k \lambda_i, \quad \quad k=1,\ldots,q,
\end{equation}
is strictly increasing.
\end{lem}

By \cite[Theorem~2]{Wit}, the optimal value of the relaxed maximization problem (\ref{max}) is
\begin{align}\label{opt}
&\frac{1}{2}\log \Big(\Big(\frac{1}{r}+\frac{1}{r}\sum_{j=1}^r \frac{1}{\Lambda_j}\Big)^r \prod_{i=1}^r \Lambda_i\Big)\nonumber\\
&=\frac{1}{2}\log\Big(\prod_{i=1}^r\Big(\frac{\Lambda_i}{r}+\frac{1}{r}\sum_{j=1}^r \frac{\Lambda_i}{\Lambda_j}\Big)\Big)
\end{align}
where $r$ is defined by Lemma~\ref{wit_l1}. Specifically, $r$
is defined by the largest eigenvalues
$\lambda_1,\lambda_2,\ldots,\lambda_q$ of $\bD_0$,
where in our case we set $q:=\min(m,N)$.

\begin{figure}
\begin{center}
\includegraphics[width=8.5cm]{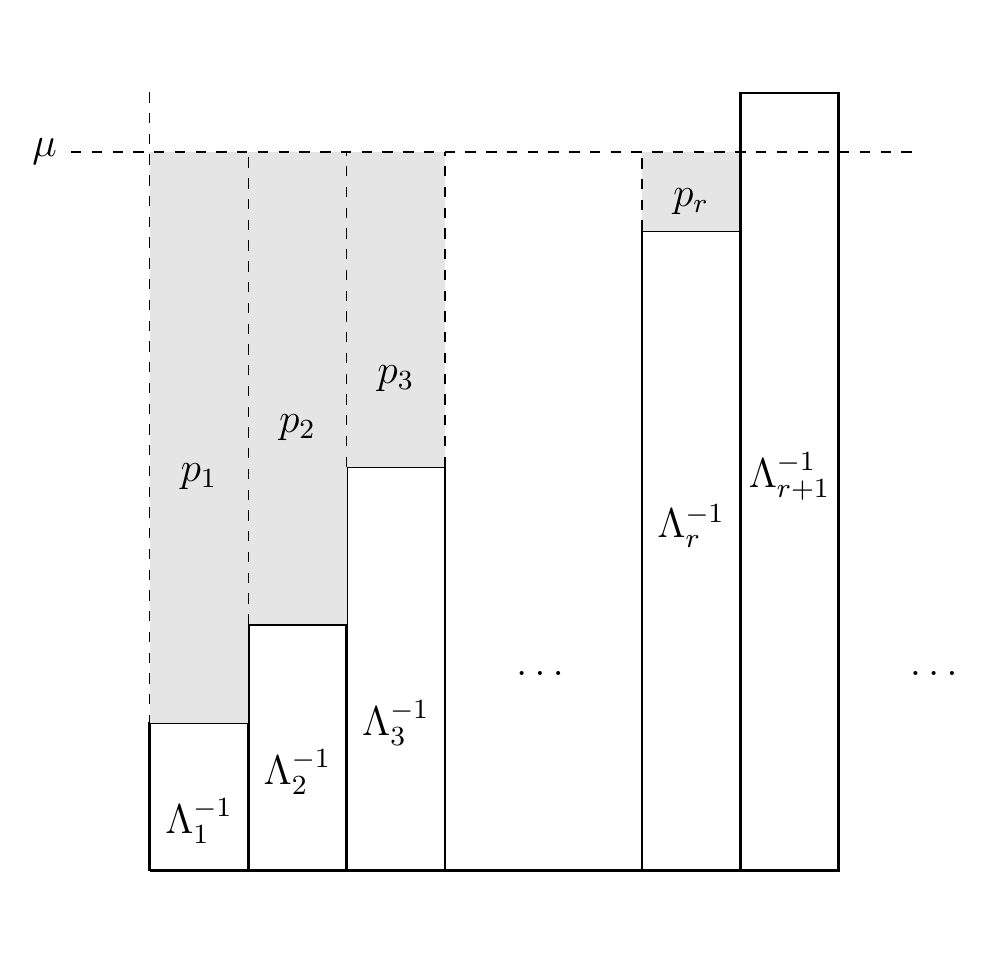}
\caption{Water-filling solution.}
\label{fig:waterfilling}
\end{center}
\end{figure}

In fact, the optimal value (\ref{opt}) may also be derived from the
solution to the well-known
water-filling problem (see \cite{Gal} for details). It is known from \cite{Wit} that
the optimal value of the maximization problem
\begin{equation}\label{prod}
\begin{split}
\text{maximize }&\prod_{i=1}^q \left(1+\Lambda_i p_i\right)\\
\text{subject to }& \sum_{i=1}^q p_i\le 1,
\end{split}
\end{equation}
is
\begin{align}
\prod_{i=1}^r\Big(\frac{\Lambda_i}{r}+\frac{1}{r}\sum_{j=1}^r \frac{\Lambda_i}{\Lambda_j}\Big).
\end{align}
This optimal value is only obtained when
\begin{equation}\label{p}
p_i=\left(\mu-\frac{1}{\Lambda_i}\right)^+,\quad i=1,2,\ldots,q,
\end{equation}
where
\begin{equation}\label{WaterLevel}
\mu:=\frac{1}{r}\Big(1+\sum_{i=1}^r \frac{1}{\Lambda_i}\Big)
\end{equation}
is called the \emph{water level}.
By taking a close look at (\ref{p}), we can see that $p_1\ge \ldots \ge p_r>0$ and $p_{r+1}=\ldots=p_q=0$.
Figure~\ref{fig:waterfilling} illustrates the relation among $\Lambda_i$,
$p_i$, and water level $\mu$.

With the values of $p_i$ defined in (\ref{p}), we can determine the $\widetilde{\bG}$ that solves the maximization problem (\ref{max}). The optimal $\widetilde{\bG}$ is
 obtained for, and only for, the following two cases. Let $\bG_0$ be the $K\times m$ matrix with
$(\bG_0)_{ii}=\sqrt{p_i}$, $i=1,\ldots,r$, and all other elements zero.
\begin{itemize}
\item Case 1. $\lambda_r>\lambda_{r+1}$ or $r=N$. Then $\widetilde{\bG}= \bG_0 \bU$ where $\bU$ is any $m\times m$ orthonormal matrix.
\item Case 2. $\lambda_i=\lambda_r$ for and only for $r-\alpha<i\le r+\beta$ with $\alpha\ge 1$, $\beta\ge1$. Then
$\widetilde{\bG}=\bdiag(\bI_{r-\alpha}, \bU_2, \bI_{K-r-\beta})  \bG_0 \bU_1$ where $\bU_1$ is any $m\times m$ orthonormal matrix and
$\bU_2$ any $(\alpha+\beta)\times(\alpha+\beta)$ orthonormal matrix.
This case is only possible when $r=q=m<N$. (The notation
$\bdiag(\bI_{r-\alpha}, \bU_2, \bI_{K-r-\beta})$ denotes a block
diagonal matrix with diagonal blocks $\bI_{r-\alpha}, \bU_2,
\bI_{K-r-\beta}$.)
\end{itemize}
After obtaining $\widetilde{\bG}$, we can extract the optimal solution $\bA=[\ba_1,\ldots,\ba_m]$ for the relaxed constraint problem (\ref{aun}) by using (\ref{gtg}), (\ref{ge}), and (\ref{ea}).

Our main motivation to relax the constraint to an \emph{average}
unit-norm constraint is our knowledge of the relaxed optimal
solution. Specifically, for the multivariate Gaussian signal $\bx$
the maximal net information gain under the relaxed constraint is
given by the water-filling solution. This helps us to identify cases
where the greedy policy is in fact optimal, as discussed in the next
section.

\section{When Greedy is Optimal}\label{g=o}

In the preceding sections, we have discussed three types of policies: the greedy policy, the optimal policy, and the relaxed optimal policy. Denote by $H_G$, $H_O$, and $H_R$ the net information gains associated with these three policies respectively. Clearly,
\begin{equation}\label{order}
H_G\le H_O\le H_R.
\end{equation}
In the rest of this section, we characterize $H_G$, $H_O$, and
$H_R$. In general, we do not expect to have $H_G=H_O$; in other
words, in general, greedy is not optimal. However, it is interesting
to explore cases where greedy \emph{is} optimal. In the rest of this
section, we provide sufficient conditions for the greedy policy to
be optimal.

Before proceeding, we make the following observation on the net information gain.
In (\ref{max}) denote $\widetilde{\bGamma}:=\widetilde{\bG}\widetilde{\bG}^T$;
then the determinant in the objective function becomes
\begin{equation}\label{det}
\det(\bI_m+\widetilde{\bG}^T\widetilde{\bLam}\widetilde{\bG})=\det(\bI_K+\widetilde{\bLam} \widetilde{\bGamma}).
\end{equation}
Under the unit-norm constraint,
\begin{align}\label{v}
\widetilde{\bGamma}&=\frac{\sigma^2}{ m} \bG\bG^T \nonumber\\
   &=\frac{\sigma^2}{ m}\Big(\sum^m_{i=1}\bg_i \bg_i^T\Big) \nonumber\\
   &=\frac{1}{ m}\bV^T\Big(\sum^m_{i=1}\ba_i \ba_i^T\Big)\bV.
\end{align}

\begin{rem}\label{ra}
In the maximization problem (\ref{un}), if the $\ba_k$s were only picked from
$\{\bv_1,\ldots,\bv_K\}$, by (\ref{v}) $\widetilde{\bGamma}=\diag(\gamma_1,\ldots,\gamma_K)$ where
  each $\gamma_i$ is an integer multiple of $1/m$ and $\sum_{k=1}^K
\gamma_k=1$. This integer $\gamma_i$ would be determined by the multiplicity of appearances of $\bv_i$ among $\ba_1,\ldots, \ba_m$. Thus the net information gain would be
\begin{equation}\label{objequal}
\frac{1}{2}\log \det(\bI_K+\widetilde{\bLam}\widetilde{\bGamma})=\frac{1}{2}\log \prod_{i=1}^K(1+\Lambda_i\gamma_i)=\frac{1}{2}\log \prod_{i=1}^N(1+\Lambda_i\gamma_i),
\end{equation}
where we use the fact that $\Lambda_{N+1}=\cdots=\Lambda_K=0$. Clearly, to maximize the net information gain by selecting compressors from $\{\bv_1,\ldots,\bv_K\}$, we should never pick $\ba_k$ from $\{\bv_{N+1},\ldots,\bv_K\}$, because (\ref{objequal})
is not a function of $\gamma_{N+1},\ldots,\gamma_K$.
In particular, the greedy policy picks $\ba_k$ from $\{\bv_1,\ldots,\bv_N\}$. After $m$ iterations of the
greedy policy, the net information gain can be computed by the right hand side of (\ref{objequal}).
\end{rem}

We now provide two sufficient conditions (in Theorems~\ref{MR2} and \ref{thm1}) under which $H_G=H_O$ holds for the sequential scalar measurements problem (\ref{ssm1}).

\begin{thm}\label{MR2}
Suppose that $\ba_k$, $k=1,\ldots,m$, can only be picked from the prescribed set
$S\subseteq\{\bv_1,\ldots,\bv_N\}$, which is a subset of the
orthonormal eigenvectors of $\bD_0$. If $\{\bv_1,\ldots,\bv_r\}\subseteq S$,
then the greedy policy is optimal, i.e., $H_G=H_O$.
\end{thm}
\begin{IEEEproof}
See Appendix \ref{App1}.
\end{IEEEproof}

Next, assume that we can pick $\ba_k$ to be any arbitrary vector with unit norm. In this much more complicated situation, we show $H_G=H_O$ by directly showing that $H_G=H_R$, which implies that $H_G=H_O$ in light of (\ref{order}).
\begin{thm}\label{thm1}
Assume that $\ba_k$, $k=1,\ldots,m$, can be selected to be any vector with $\|\ba_k\|\leq 1$.
If $1/\lambda_{k}-1/\lambda_{k+1}=n_k/\sigma^{2}$, where $n_k$ is
some nonnegative integer, for $k=1,\ldots,r-1$, and $r$ divides
$(m-\sum_{k=1}^{r-1} kn_k)$, then the greedy policy is optimal,
i.e., $H_G=H_O$
\end{thm}
\begin{IEEEproof}
See Appendix \ref{app2}
\end{IEEEproof}

The two theorems above furnish conditions under which greedy is optimal.
However, these conditions are quite restrictive. Indeed, as pointed
out earlier, in general the greedy policy is not optimal. The
restrictiveness of the sufficient conditions above help to highlight
this fact. In the next section, we provide examples of cases where
greedy is \emph{not} optimal.

\section{When Greedy Is Not Optimal}\label{examples}

\subsection{An Example with Non-Scalar Measurements}
In this subsection we give an example where the greedy policy is not
optimal for the scenario $\bz=\bx$ and $\by_k=\bA_k \bx+\bw_k$.
Suppose that we are restricted to a set of only three choices for $\bA_k$:
\[
\Ac=\left\{\diag(1,0),\diag(0,1),\frac{1}{2}\diag(1,1)\right\}.
\]
Note that $\diag(1,1)=\bI$.
In this case, $L=N=2$. Moreover, set $m=2$,
$\bP_0=16\bI$, and $\bR_{ww}=\bI$.

Let us see what the greedy policy would do in
this case. For $k=1$, it would pick $\bA_1$ to maximize
\[
\det\left(\frac{1}{16}\bI + (\bA_1)^2\right).
\]
A quick calculation shows that
for $\bA_1=\diag(1,0)$ or $\diag(0,1)$, we have
\[
\det\left(\frac{1}{16}\bI + (\bA_1)^2\right) = \frac{17}{256},
\]
whereas for $\bA_1 = \frac{1}{2}\diag(1,1)$,
\[
\det\left(\frac{1}{16}\bI + (\bA_1)^2\right) =  \frac{25}{256},
\]
So the greedy policy picks $\bA_1 = \frac{1}{2}\diag(1,1)$,
which leads to $\bP_1 = \frac{16}{5}\bI$.

For $k=2$, we go through the same
calculations: for $\bA_2=\diag(1,0)$ or $\diag(0,1)$, we have
\[
\det\left(\frac{5}{16}\bI + (\bA_2)^2\right) = \frac{105}{256}
\]
whereas for $\bA_2 = \frac{1}{2}\diag(1,1)$,
\[
\det\left(\frac{5}{16}\bI + (\bA_2)^2\right) = \frac{81}{256}.
\]
So, this time the greedy policy picks $\bA_2 = \diag(1,0)$ (or
$\diag(0,1)$), after which $\det(\bP_2)=256/105$.

Consider the alternative policy that picks $\bA_1=\diag(1,0)$ and
$\bA_2=\diag(0,1)$. In this case,
\begin{align}
\bP_2^{-1} = \frac{1}{16}\bI + \diag(1,0) + \diag(0,1)= \frac{17}{16}\bI
\end{align}
and so $\det(\bP_2) = 256/289$, which is clearly provides greater net information gain than the greedy policy. Call this alternative policy the
\emph{alternating policy} (because it alternates between
$\diag(1,0)$ and $\diag(0,1)$).

In conclusion, for this example the greedy policy
is not optimal with respect to the objective of maximizing the
net information gain. How much worse is the objective function of the greedy policy relative to that of the optimal policy? On the face of it, this question seems easy to answer in light of the
well-known fact that the net information gain is a submodular function.
As mentioned before, in this case we would expect to be able to bound the suboptimality of the greedy policy compared to the optimal policy
(though we do not explicitly do that here).

Nonetheless, it is worthwhile exploring this question a little further.
Suppose that we set $\bP_0 = \alpha^{-1}\bI$ and let the third choice in
$\Ac$ be $\alpha^{1/4}\bI$, where $\alpha>0$ is some small number.
(Note that the numerical example above is a special case with $\alpha=1/16$.)
In this case, it is straightforward to check that
the greedy policy picks $\bA_1=\alpha^{1/4}\bI$ and
$\bA_2=\diag(1,0)$ (or $\diag(0,1)$) if $\alpha$ is sufficiently
small, resulting in
\[
\det(\bP_2) =
\frac{1}{\sqrt{\alpha}(1+\sqrt{\alpha})(1+\sqrt{\alpha}+\alpha)},
\]
which increases unboundedly as $\alpha\to 0$.
However, the alternating policy results in
\[
\det(\bP_2) = \frac{1}{(1+\alpha)^2},
\]
which converges to $1$ as $\alpha\to 0$. Hence, letting $\alpha$ get arbitrarily small, the ratio of $\det(\bP_2)$ for the greedy policy to that of the alternating policy can be made arbitrarily large. Insofar as we accept minimizing $\det(\bP_2)$ to be an equivalent objective to maximizing the net information gain (which differs by the normalizing factor $\det(\bP_0)$ and taking $\log$), this means that \emph{the greedy policy is arbitrarily worse than the alternating policy}.

What went wrong? The greedy policy was ``fooled'' into picking
$\bA_1 =\alpha^{1/4}\bI$ at the first stage,
because this choice maximizes the per-stage information gain in the
first stage. But once it does that, it is stuck with its resulting
covariance matrix $\bP_1$. The alternating policy trades off the per-stage
information gain in the first stage for the sake of better net
information gain over two stages. The first measurement matrix
$\diag(1,0)$ ``sets up'' the covariance matrix $\bP_1$ so that the
second measurement matrix $\diag(0,1)$ can take advantage of it to
obtain a superior covariance matrix $\bP_2$ after the second stage,
embodying a form of ``delayed gratification.''

Interestingly, the argument above depends on the value of $\alpha$
being sufficiently small. For example, if $\alpha=0.347809$, then the
greedy policy has the same net information gain as the alternating
policy, and is in fact optimal.

An interesting observation to be made here is that the submodularity of the net information gain as an objective function depends crucially on including the $\log$ function. In other words, although for the purpose of optimization we can dispense with the $\log$ function in the objective function in view of its monotonicity, bounding the suboptimality of the greedy policy with respect to the optimal policy turns on submodularity, which relies on the presence of the $\log$ function in the objective function. In particular, if we adopt the volume of the error concentration ellipse as an equivalent objective function, we can no longer bound the suboptimality of the greedy policy relative to the optimal policy---the greedy policy is provably \emph{arbitrarily worse} in some scenarios, as our example above shows.

\subsection{An Example with Scalar Measurements}

Consider the channel model $\bz=\bx$ and scalar measurements $y_k=\ba^T_k\bx+w_k$. Assume that
\[
\bP_0=\left[ \begin{array}{cc}
                3 & 2 \\
                2 & 3
              \end{array}
\right],
\]
$\bR_{ww}=\bI$, and set $m=2$. Our goal is to find
$\|\ba_1\|,\|\ba_2\|\le 1$ such that $\ba_1$, $\ba_2$ maximize the net information gain:
\begin{equation}
H_0-H_2=\frac{1}{2}\log \det(\bP_0)\det(\bP_0^{-1}+\ba_1\ba_1^T+\ba_2\ba_2^T).
\end{equation}
By simple computation, we know that the eigenvalues of $\bP_0$ are $\lambda^{(0)}_1=5$ and $\lambda^{(0)}_2=1$. If we follow the greedy
policy, the eigenvalues of $\bP_1$ are $\lambda_1^{(1)}=1$ and $\lambda_2^{(1)}=5/6$. By (\ref{greedy}), the net information
gain for the greedy policy is
$$
H_0-H_2=\frac{1}{2}\log(1+5)(1+1)=\frac{1}{2}\log(12).
$$

Next we solve for the optimal solution. Let $\ba_1=[a_1,a_2]^T$. By (\ref{PP}), we have
\begin{equation*}
\bP_1=\left[\begin{array}{cc}
            \frac{5a_2^2 + 3}{3a_1^2 + 4a_1a_2 + 3a_2^2 + 1}    & \frac{-(5a_1a_2 - 2)}{3a_1^2 + 4a_1a_2 + 3a_2^2 + 1} \\
           \frac{-(5a_1a_2 - 2)}{3a_1^2 + 4a_1a_2 + 3a_2^2 + 1}   &  \frac{5a_1^2 + 3}{3a_1^2 + 4a_1a_2 + 3a_2^2 + 1}
          \end{array}
\right].
\end{equation*}
We compute that
\begin{align}
\lambda_1^{(1)}
&=\frac{(25a_1^4 + 50a_1^2a_2^2 - 80a_1a_2 + 25a_2^4 + 16)^{1/2}}{6a_1^2 + 8a_1a_2 + 6a_2^2 + 2}\nonumber\\
 &\qquad + \frac{5a_1^2 + 5a_2^2 + 6}{6a_1^2 + 8a_1a_2 + 6a_2^2 + 2}.
\end{align}
When we choose $\ba_2$ in the second stage, we can simply maximize the information gain in that stage. In this special case when $m=2$, the second stage is actually the last one.
If $\ba_1$ is given, maximizing the net information gain is equivalent to maximizing the information gain in the second stage.
Therefore, the second step is equivalent to a greedy step. By (\ref{greedy}),
\begin{align}\label{i1}
H_1-H_2&=-\frac{1}{2}\log\left(1-\frac{1}{1+1/\lambda_1^{(1)}}\right)\nonumber\\
&=\frac{1}{2}\log(1+\lambda^{(1)}_1).
\end{align}
By (\ref{infogain}), we know
\begin{align}\label{i2}
H_0-H_1&=-\frac{1}{2}\log\det\left(\bI_2-\frac{\bP_0 \ba_1\ba_1^T}{\ba_1^T\bP_0 \ba_1+1}\right)\nonumber\\
&= \frac{1}{2}\log\left(4+4a_1a_2\right).
\end{align}
Using $\|\ba_1\|=1$, we simplify (\ref{i1}) and (\ref{i2}) to obtain
\begin{align}
\MoveEqLeft H_0-H_2=\frac{1}{2}\log\Big(\frac{1}{2}\big((41-80a_1a_2)^{1/2}\nonumber\\
&+19+8a_1a_2\big)\Big).
\end{align}
This expression reaches its maximal value when $a_1a_2=1/5$. So the optimal net information gain is
$\frac{1}{2}\log(12.8)$, when
\begin{equation*}
\ba_1=\left[\left(\frac{-\sqrt{21}+5}{10}\right)^{1/2},\left(\frac{\sqrt{21}+5}{10}\right)^{1/2}\right]^T
\end{equation*}
and
\begin{equation*}
\ba_2=\left[\left(\frac{\sqrt{21}+5}{10}\right)^{1/2},\left(\frac{-\sqrt{21}+5}{10}\right)^{1/2}\right]^T.
\end{equation*}
This implies that the greedy policy is not optimal.

\appendices
\section{Proof of Theorem 4}\label{App1}

If $\ba_k$, $k=1,\ldots,m$, can only be picked from
$\{\bv_1,\ldots,\bv_N\}$, then by (\ref{objequal}) the net information gain is $\frac{1}{2}\log \prod_{i=1}^N(1+\Lambda_i\gamma_i)$. We can simply manage $\gamma_i$ in each channel to maximize
the net information gain.
Rewrite
\begin{equation}\label{heights}
\prod_{k=1}^N (1+\Lambda_k \gamma_k)=\big(\prod_{k=1}^N \Lambda_k\big) \prod_{k=1}^N \big(\frac{1}{\Lambda_k}+\gamma_k\big).
\end{equation}
As we claimed before, $\sum_{i=1}^N \gamma_k=1$ where $\gamma_k$,
$k=1,\ldots,N$, is an integer multiple of $1/m$. Inspired by the
water-filling algorithm, we can consider
$(\gamma_1,\ldots,\gamma_N)$ as an allocation of $m$ blocks (each
with size $1/m$) into $N$ channels. In contrast to water-filling, we
refer to this problem as \emph{block-filling} (or, to be more
evocative, \emph{ice-cube-filling}).
The original heights of these channels are $1/\Lambda_1\le \ldots\le 1/\Lambda_N$. Finally, the net information gain is determined by the product $\prod_{k=1}^N(\frac{1}{\Lambda_k}+\gamma_k)$ of the final heights. The optimal solution can be extracted from an optimal allocation that maximizes (\ref{heights}).

Because $\Lambda_1\ge \ldots\ge \Lambda_N$, to maximize
$\prod_{k=1}^N (1+\Lambda_k \gamma_k)$ we should allocate nonzero values of $\gamma_k$ in the first
$q=\min(m,N)$ channels.
Accordingly, there exists an optimal solution $\boldsymbol{\alpha}=(\alpha_1,\ldots,\alpha_N)$ such that
\begin{equation}\label{heights2}
\prod_{k=1}^N (1+\Lambda_k \alpha_k)= \prod_{k=1}^q (1+{\Lambda_k}\alpha_k).
\end{equation}

Assume that we pick $\ba_k$, $k=1,\ldots,m$, using the greedy policy.
By (\ref{v1}) and (\ref{v2}), we see that the $k$th iteration of the
greedy algorithm only changes $\Lambda^{(k-1)}_1$ into
$(\frac{1}{\Lambda^{(k-1)}_1}+\frac{1}{m})^{-1}$, which is equivalent to changing $\frac{1}{\Lambda^{(k-1)}_1}$ into $\frac{1}{\Lambda^{(k-1)}_1}+\frac{1}{m}$. Consider this greedy policy
in the viewpoint of block-filling. The greedy policy fills blocks to the lowest channel one by one. If there are more than one channel having the same lowest height, it adds to the
channel with the smallest index.
Likewise, since the original heights of the channels are $1/\Lambda_1\le \ldots\le 1/\Lambda_N$,
the greedy policy only fills blocks to the first $q$ channels, i.e., greedy
solution $\boldsymbol{\eta}=(\eta_1,\ldots,\eta_N)$ also satisfies
\begin{equation}\label{heights3}
\prod_{k=1}^N (1+\Lambda_k \eta_k)=\prod_{k=1}^q  (1+{\Lambda_k}\eta_k ).
\end{equation}

 We now provide a necessary condition for both optimal and greedy solutions.


\begin{lem}\label{necess}
Assume that an allocation $\boldsymbol{\alpha}$ is determined by either an optimal solution or a greedy solution. If $\alpha_k$ is nonzero, then $\alpha_k+\frac{1}{\Lambda_k}$ is bounded in the interval
$(\mu-\frac{1}{m},\mu+\frac{1}{m})$. Moreover, it suffices for the optimal and greedy solutions to pick from the set  $\{\bv_1,\ldots,\bv_r\}$.

\end{lem}
\begin{IEEEproof}
 First, assume that $\boldsymbol{\alpha}$ is given by an optimal
solution. Recall that $\alpha_k+\frac{1}{\Lambda_k}$ is the final
height of the $k$th channel. By examining the total volumes of water
and blocks, we deduce the following. If $\alpha_i>0$ and
$\alpha_i+\frac{1}{\Lambda_i}> \mu$ for some $1\le i\le q$, where
$\mu$ is the water level defined in (\ref{WaterLevel}), then there
exists some channel $1\le j\le r$ such that
$\alpha_j+\frac{1}{\Lambda_j}<\mu$. For the purpose of proof by contradiction, let us assume that $\alpha_i+\frac{1}{\Lambda_i}\ge \mu+\frac{1}{m}$. We move the top block of the $i$th channel to the $j$th channel to get another allocation $\boldsymbol{\beta}=(\beta_1,\ldots,\beta_m)$. Clearly, $\boldsymbol{\beta}$ and $\boldsymbol{\alpha}$ have the same entries except the $i$th and $j$th components. The argument in this paragraph is illustrated in Figure~\ref{fig:etagamma}.

\begin{figure}
\begin{center}
\includegraphics[width=8.5cm]{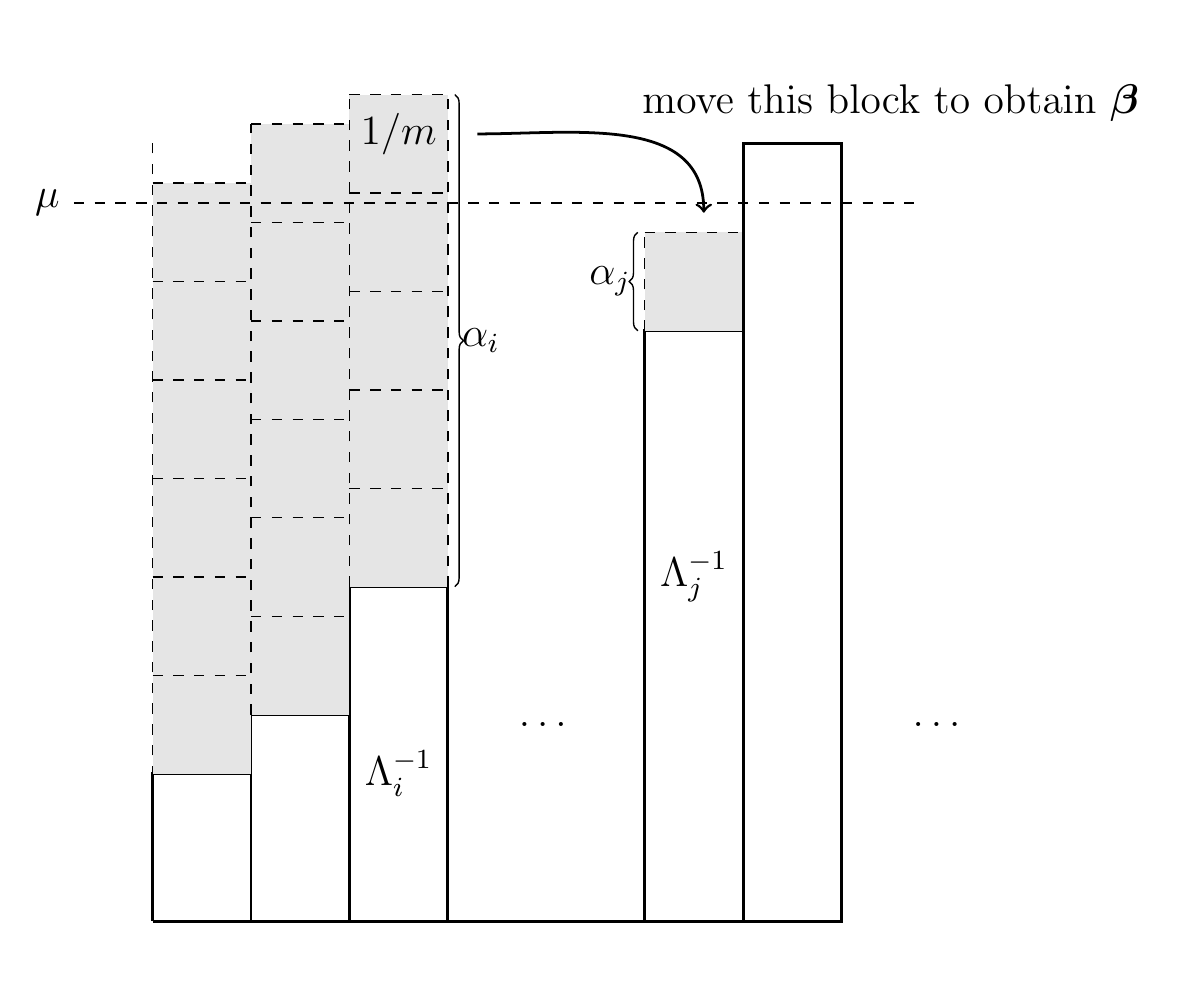}
\caption{Obtain allocation $\boldsymbol{\mathbf{\beta}}$ from $\boldsymbol{\alpha}$.}
\label{fig:etagamma}
\end{center}
\end{figure}

 For simplicity, denote $\delta_k:=\alpha_k+\Lambda_k^{-1}-\mu$ for $k=1,\ldots,m$. So
   \begin{align}\label{move}
   &\frac{\prod_{k=1}^m(1+\Lambda_k \beta_k)}{\prod_{i=k}^m(1+\Lambda_k \alpha_k)}\nonumber\\
   &=\frac{(1+\Lambda_i(\mu+\delta_i-\Lambda_i^{-1}-m^{-1}))}{(1+\Lambda_i(\mu +\delta_i-\Lambda_i^{-1}))}\nonumber\\
   &~~\frac{(1+\Lambda_j(\mu+\delta_j-\Lambda_j^{-1}+m^{-1}))}{(1+\Lambda_i(\mu +\delta_j-\Lambda_j^{-1}))}\nonumber\\
   &=\frac{( \mu +\delta_i-m^{-1})(\mu +\delta_j+m^{-1})}{(\mu +\delta_i)( \mu+\delta_j)}\nonumber\\
   &=\frac{(\mu+\delta_i)(\mu+\delta_j)+m^{-1}(\delta_i-\delta_j)-m^{-2}}{(\mu+\delta_i)(\mu+\delta_j)}>1,
   \end{align}
    because $\delta_i-\delta_j>\frac{1}{m}$. Thus $\boldsymbol{\beta}$ gives a better allocation, which contradicts the optimality of $\boldsymbol{\alpha}$. By a similar argument, we obtain that
   for any optimal solution $\boldsymbol{\alpha}$, there also does not exist $i$ such that $\alpha_i>0$ and  $\alpha_i+\frac{1}{\Lambda_i}\le \mu-\frac{1}{m}$.
   In conclusion, the final height $\alpha_i+\frac{1}{\Lambda_i}$, $i=1,\ldots,r$, in each channel in the optimal solution is bounded in the interval $(\mu-\frac{1}{m},\mu+\frac{1}{m})$.
   Additionally, in both cases when $r=q$ and $r<q$, $\alpha_{r+1}=\cdots=\alpha_N=0$. This means that it suffices for the optimal solution to pick from the set  $\{\bv_1,\ldots,\bv_r\}$.

Next, we assume that $\boldsymbol{\alpha}$ is determined by a greedy solution.
If $\alpha_i>0$ and $\alpha_i+\frac{1}{\Lambda_i}> \mu$, for some $1\le i\le q$, then there exists a channel
with index $1\le j\le r$ such that
$\eta_j+\frac{1}{\Lambda_j}<\mu$. For the purpose of proof by
contradiction, let us assume that $\alpha_i+\frac{1}{\Lambda_i}\ge
\mu+\frac{1}{m}$. This implies that when the greedy algorithm fills
the top block to the  $i$th channel, it does not add that block to
the $j$th channel with a lower height. This contradicts how the
the greedy policy actually behaves. By a similar argument, there
does not exist some channel $i$ such that $\alpha_i>0$ and
$\alpha_i+\frac{1}{\Lambda_i}\le \mu-\frac{1}{m}$. In conclusion,
the final height $\alpha_i+\frac{1}{\Lambda_i}$, $i=1,\ldots,r$, in
each channel in the greedy solution is bounded in the interval
$(\mu-\frac{1}{m},\mu+\frac{1}{m})$. Moreover,
$\alpha_{r+1}=\cdots=\alpha_N=0$. This means that it suffices for the greedy solution to pick from the set  $\{\bv_1,\ldots,\bv_r\}$.
\end{IEEEproof}

We now proceed to the equivalence between the optimal solution and the greedy solution.
To show this equivalence, let $\boldsymbol{\theta} =
(\theta_1,\ldots,\theta_r)$ be an arbitrary allocation of $m$ blocks
satisfying the necessary condition in Lemma~\ref{necess}.
Next, we will show how to modify $\boldsymbol{\theta}$ to obtain an
optimal allocation. After that, we will also show how to modify $\boldsymbol{\theta}$ to obtain an allocation that is generated by the greedy policy. It will then be evident that these two resulting allocations have the same information gain.

To obtain an optimal allocation from $\boldsymbol{\theta}$, we first remove the top block from each channel whose height is above $\mu$ to get an auxiliary allocation $\boldsymbol{\theta}'=(\theta'_1,\ldots,\theta'_r)$. Assume that the total number of removed blocks is $m'$. This auxiliary $\boldsymbol{\theta}'$ is unique, because each $\theta'_k$ is simply the maximal number of blocks can be filled in the $k$th channel to obtain a height not above the water level: this number is uniquely determined by $\Lambda_k$, $\mu$, and $m$. We now show how to re-allocate the removed $m'$ blocks, so that, together with $\boldsymbol{\theta}'$, we have an optimal allocation of all $m$ blocks.

Note that by Lemma~\ref{necess}, to obtain an optimal solution we
cannot allocate more than one block to any channel, because that
would make the height of that channel above $\mu+\frac{1}{m}$. We
claim that the optimal allocation simply re-allocates the $m'$ removed blocks to the lowest $m'$ channels in $\boldsymbol{\theta}'$. We can show this by contradiction.
Assume that the optimal allocation adds one block to the $i$th channel instead of a lower $j$th channel in $\boldsymbol{\theta}'$. This means that $\theta'_i>\theta'_j$, $\theta_i=\theta'_i+1/m$, and $\theta_j=\theta'_j$. By an argument similar to (\ref{move}),
if we move the top block in the $i$th channel to the $j$th channel, we would obtain a better allocation (which gives a larger net information gain). This contradiction verifies our claim.

Next, we concentrate on the allocation provided by the greedy
policy. First, we recall that at each step of the greedy algorithm it
never fills a block to some higher channel instead of a lower one.
So after the greedy algorithm fills one block to some channel, its
height cannot differ from a lower channel by more than $1/m$. If we
apply the greedy policy for picking $\ba_k$, $k=1,\ldots,(m-m')$,
then we obtain the same allocation as $\boldsymbol{\theta}'$. This
is because any other allocation of $(m-m')$ blocks would result in a
channel, after its top block filled, with a height deviating by more than $1/m$ from some other channel. This allocation contradicts the behavior of the greedy policy. Continuing with $\boldsymbol{\theta}'$, the greedy policy simply allocates the remaining $m'$ blocks to the lowest $m'$ channels one by one. So the greedy policy gives the same final heights as the optimal allocation. The only possible difference is the order of these heights. Therefore, the greedy solution is equivalent to the optimal solution in the sense of giving the same net information gain, i.e., $H_G=H_O$. This completes the proof of Theorem \ref{MR2}.

\section{Proof of Theorem 5}\label{app2}
\begin{figure}
\begin{center}
\includegraphics[width=8.5cm]{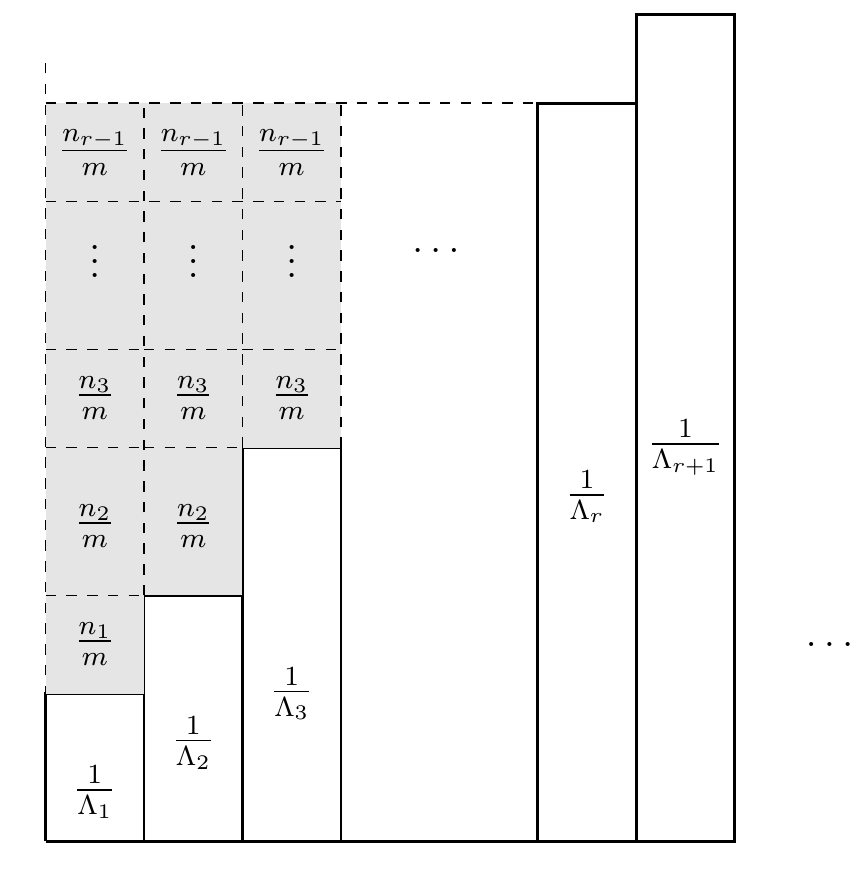}
\caption{Heights of channels after $\hat{m}$ iterations of the greedy policy}
\label{fig:integer}
\end{center}
\end{figure}

We have studied the performance of the greedy policy in the viewpoint of block-filling in the proof of Theorem \ref{MR2}.
For the purpose of simplicity, we rewrite $1/\lambda_{k}-1/\lambda_{k+1}=n_k/\sigma^{2}$ as
\begin{equation}\label{Lam}
\frac{1}{\Lambda_{k}}-\frac{1}{\Lambda_{k+1}}=\frac{n_k}{m}
\end{equation}
where $\Lambda_i=\frac{m\lambda_i}{\sigma^2}$.
After $\hat{m}:=\sum_{k=1}^{r-1} kn_k$ iterations of the greedy policy, the heights in the first $r$ channels give a flat top, which is illustrated in Figure \ref{fig:integer}.

There are $m-\hat{m}$ blocks remaining after $\hat{m}$ iterations. If $r$ divides $m-\hat{m}$, the final heights of the first $r$ channels still
give a flat top coinciding with $\mu$ in each channel. Therefore $H_G=H_R$. From (\ref{order}),
we conclude that $H_G=H_O$.

%








\begin{thebibliography}{9}
\bibitem{AHN}A. Ashok, J. L. Huang, and M. A. Neifeld, ``Information-optimal Adaptive Compressive Imaging,'' {\it Proc. of the Asilomar Conf. on Signals, Systems, and Computers},
Pacific Grove, CA, Nov. 2011, pp.~1255--1259.
\bibitem{BV} S. Boyd and L. Vandenberghe, \emph{Convex Optimization}. Cambridge, MA: Cambridge University Press, 2004.
\bibitem{CCPV} G. Calinescu, C. Chekuri, M. Pal, and J. Vondrak, ``Maximizing a
monotone submodular function subject to a matroid constraint,'' {\it the 20th SICOMP Conf.}, 2009.
\bibitem{Car}  W. R. Carson, M. Chen, M. R. D. Rodrigues,
R. Calderbank, and L. Carin, ``Communications-Inspired Projection Design with Application to Compressive Sensing,'' Preprint.
\bibitem{CaH08}R. Castro, J. Haupt, R. Nowak, and G. Raz, ``Finding needles in noisy haystacks,'' {\it Proc.\ IEEE Intl.\ Conf.\ on Acoustics, Speech and Signal Processing}, Las Vegas, NV, Apr. 2008, pp.~5133--5136.
\bibitem{DZ}J. Ding and A. Zhou, ``Eigenvalues of rank-one updated matrices with some applications,'' {\it Applied Mathematics Letters}, vol.~20, no.~12, pp.~1223--1226, 2007.
\bibitem{Don}D. L. Donoho, ``Compressed sensing,'' {\it IEEE Trans. Inf. Theory}, vol.~52, no.~4, pp.~1289--1306, 2006.
\bibitem{Ela07} M. Elad, ``Optimized projections for compressed sensing,'' {\it IEEE Trans. Signal Process.}, vol.~55, no.~12, pp.~5695--5702, 2007.
\bibitem{Gal} R. G. Gallager, \emph{Information Theory and Reliable Communication}. New York: John Wiley \& Sons, Inc., 1968.
\bibitem{HaC10} J. Haupt, R. Castro, and R. Nowak, ``Distilled sensing: Adaptive sampling for sparse detection and estimation,'' preprint, Jan. 2010 [online]. Available:
http://www.ece.umn.edu/$\sim$jdhaupt/publications/sub10\_ds.pdf
\bibitem{HCN10}J. Haupt, R. Castro, and R. Nowak, ``Improved bounds for sparse recovery from adaptive measurements,'' {\it ISIT 2010}, Austin, TX, Jun. 2010.
\bibitem{HJ}R. A. Horn and C. R. Johnson, \emph{Matrix Analysis}. Cambridge, MA: Cambridge University Press, 1985.
\bibitem{JiD09} S. Ji, D. Dunson, and L. Carin, ``Multitask compressive sensing,'' {\it IEEE Trans. Signal Process.} vol.~57, no.~1, pp.~92--106, 2009.
\bibitem{JXL} S. Ji, Y. Xue and L. Carin, ``Bayesian compressive sensing,'' {\it IEEE Trans. Signal Process.}, vol.~56, no.~6, pp.~2346--2356, 2008.
\bibitem{JoBo}S. Joshi and S. Boyd, ``Sensor selection via convex optimization,'' {\it IEEE Trans. Signal Process.}, vol.~57, no.~2, pp.~451--462, 2009.
\bibitem{KAN} J. Ke, A. Ashok, and M. A. Neifeld, ``Object reconstruction from adaptive compressive
measurements in feature-specific imaging'', {\it Applied Optics}, vol.~49, no.~34, pp.~H27-H39, 2010.
\bibitem{LC} E. Liu and E. K. P. Chong, ``On Greedy Adaptive Measurements,'' {\it Proc. CISS}, 2012.
\bibitem{LCS} E. Liu, E. K. P. Chong, and L. L. Scharf ``Greedy Adaptive Measurements with Signal and Measurement Noise,'' submitted to Asilomar conf. on signals, systems, and Computers, Mar. 2012.
\bibitem{NW}G. L. Nemhauser and L. A. Wolsey, ``Best algorithms for approximating the maximum of a submodular set function,'' {\it Math. Oper. Research}, vol.~3, no.~3, pp.~177--188, 1978.
\bibitem{PRV} F. P\'{e}rez-Cruz, M. R. Rodrigues, and S. Verd\'{u}, ``MIMO Gaussian channels with arbitrary inputs: Optimal precoding and power allocation,'' {\it IEEE Trans. Inf. Theory}, vol.~56,
no.~3, pp.~1070--1084, 2010.
\bibitem{REJVBBP}H. Rowaihy, S. Eswaran, M. Johnson, D. Verma, A. Bar-Noy, T. Brown, and T. L. Portal, ``A survey of sensor selection schemes
in wireless sensor networks,'' {\it Proc.\ SPIE}, 2007, vol.~6562.
\bibitem{SBV}M. Shamaiah, S. Banerjee and H. Vikalo, ``Greedy sensor selection: Leveraging submodularity,'' {\it Proc.\ of the 49th IEEE Conf. on Decision and Control}, Atlanta, GA, Dec. 2010.
\bibitem{WPR}D. P. Wipf, J. A. Palmer, and B. D. Rao, ``Perspectives on sparse Bayesian learning,'' {\it Neural Information Processing Systems (NIPS)}, Vancouver, Canada, Dec. 2004.
\bibitem{Wit} H. S. Witsenhausen, ``A determinant maximization problem occurring in the theory of data communication,''
{\it SIAM J.\ Appl.\ Math}, vol.~29, no.~3, pp.~515--522, 1975.
\end{thebibliography}
\end{document}